\documentclass[a4paper,fleqn,usenatbib]{mnras}   

\usepackage[T1]{fontenc}
\usepackage{ae,aecompl}
\usepackage{graphicx}	
\usepackage{amsmath}	
\usepackage{amssymb}	

\newcommand{\hii}{\textrm{H}\textsc{ii}}
\newcommand{\hi}{\textrm{H}\textsc{i}}

\newcommand{\oiidoub}{[\textrm{O}\textsc{ii}]\ensuremath{\lambda3727,3729}}

\newcommand{\oiiiv}{[\textrm{O}\textsc{iii}]\ensuremath{\lambda5007}}
\newcommand{\oiiiiv}{[\textrm{O}\textsc{iii}]\ensuremath{\lambda4959}}

\newcommand{\ha}{\ifmmode {\rm H}\alpha \else H$\alpha$\fi}
\newcommand{\hb}{\ifmmode {\rm H}\beta \else H$\beta$\fi}
\newcommand{\lya}{\ifmmode {\rm Ly}\alpha \else Ly$\alpha$\fi}
\newcommand{\pg}{\ifmmode {\rm P}\gamma \else Pa$\gamma$\fi}
\newcommand{\lyb}{\ifmmode {\rm Ly}\beta \else Ly$\beta$\fi}
\newcommand{\lyg}{\ifmmode {\rm Ly}\gamma \else Ly$\gamma$\fi}

\newcommand{\ciiidoub}{\textrm{C}\textsc{iii}]\ensuremath{\lambda\lambda1907,1909}}

\newcommand{\civ}{\textrm{C}\textsc{iv}\ensuremath{\lambda1548,1550}}

\newcommand{\heii}{\textrm{He}\textsc{ii}\ensuremath{\lambda1640}}
\newcommand{\oiiiuv}{\textrm{O}\textsc{iii}]\ensuremath{\lambda1661,1666}}

\newcommand{\flyc}{\ifmmode  \mathrm{f}_\mathrm{esc}\mathrm{(LyC)} \else $\mathrm{f}_\mathrm{esc}\mathrm{(LyC)}$\fi}

\def\kms{km s$^{-1}$}

\def\ergs{\ifmmode \mathrm{erg\hspace{1mm}s}^{-1} \else erg s$^{-1}$\fi}
\def\ergscm{erg s$^{-1}$ cm$^{-2}$}
\def\micron{\ifmmode \mu\mathrm{m} \else $\mu$m\fi}
\def\msun{\ifmmode \mathrm{M}_{\odot} \else M$_{\odot}$\fi}
\def\msunyr{\ifmmode \mathrm{M}_{\odot} \hspace{1mm}{\rm yr}^{-1} \else $\mathrm{M}_{\odot}$ yr$^{-1}$\fi}
\def\zsun{\ifmmode Z_{\odot} \else Z$_{\odot}$\fi}
\def\lsun{\ifmmode L_{\odot} \else L$_{\odot}$\fi}
\def\mstar{\ifmmode \mathrm{M}_{\star} \else M$_{\star}$\fi}

\title[\lya\ emission crossing the caustic]
{Candidate Population III stellar complex at z=6.629 in the MUSE Deep Lensed Field}

\author[E.~Vanzella et al.]{
\parbox[t]{\textwidth}{E.~Vanzella$^1$\thanks{E-mail: eros.vanzella@inaf.it},
M. Meneghetti$^1$, G. B. Caminha$^2$, M. Castellano$^3$, F. Calura$^1$, P. Rosati$^{4,1}$,
C. Grillo$^5$, M. Dijkstra, M. Gronke$^6$, E. Sani$^7$, A. Mercurio$^8$, P. Tozzi$^9$, M. Nonino$^{10}$, S. Cristiani$^{10}$, 
M. Mignoli$^1$, L. Pentericci$^3$, R. Gilli$^1$, T. Treu$^{11}$, K. Caputi$^2$, G. Cupani$^{10}$, A. Fontana$^3$,  A. Grazian$^{12}$ 
and I. Balestra$^{10,13}$
}
\vspace*{8pt}\\
$^1$INAF -- Osservatorio di Astrofisica e Scienza dello Spazio, via Gobetti 93/3, 40129 Bologna, Italy\\
$^2$Kapteyn Astronomical Institute, University of Groningen, Postbus 800, 9700 AV Groningen, The Netherlands\\
$^3$INAF -- Osservatorio Astronomico di Roma, Via Frascati 33, I-00078 Monte Porzio Catone (RM), Italy\\
$^4$Dipartimento di Fisica e Scienze della Terra, Universit\`a degli Studi di Ferrara, via Saragat 1, I-44122 Ferrara, Italy\\
$^5$Dipartimento di Fisica, Universit\`a  degli Studi di Milano, via Celoria 16, I-20133 Milano, Italy\\
$^6$Department of Physics, University of California, Santa Barbara, CA 93106, USA\\
$^7$European Southern Observatory, Alonso de Cordova 3107, Casilla 19, Santiago 19001, Chile \\
$^8$INAF -- Osservatorio Astronomico di Capodimonte, Via Moiariello 16, I-80131 Napoli, Italy\\
$^9$INAF -- Osservatorio Astrofisico di Arcetri, Largo E. Fermi, I-50125, Firenze, Italy\\
$^{10}$INAF -- Osservatorio Astronomico di Trieste, via G. B. Tiepolo 11, I-34143, Trieste, Italy\\
$^{11}$Department of Physics and Astronomy, University of California, Los Angeles, CA 90095, USA\\
$^{12}$INAF -- Osservatorio Astronomico di Padova, Vicolo Osservatorio 5, 35122, Padova, Italy\\
$^{13}$OmegaLambdaTec GmbH, Lichtenbergstrasse 8, 85748 Garching bei Munchen, Germany
}

\begin{document}
\date{}
\maketitle

\begin{abstract}
We discovered a strongly lensed ($\mu \gtrsim 40$) \lya\ emission at z=6.629 (S/N $\simeq 18$) in
the MUSE Deep Lensed Field (MDLF) targeting the Hubble Frontier Field galaxy cluster MACS~J0416. 
Dedicated lensing simulations imply that the \lya\ emitting region necessarily crosses the caustic.
The arc-like shape of the \lya\ extends 3$''$ on the observed plane and is the result of two merged 
multiple images, each one with a de-lensed \lya\ luminosity L $\lesssim 2.8\times10^{40}$ \ergs\ arising 
from a confined region ($\lesssim 150$ pc effective radius). A spatially unresolved HST counterpart is 
barely detected at S/N $\simeq2$ after stacking the  near-infrared bands, corresponding to an 
observed(intrinsic) magnitude $m_{1500} \gtrsim 30.8(\gtrsim35.0)$. The inferred rest-frame \lya\ 
equivalent width is EW$_{\rm 0}$ > 1120\AA~if the IGM transmission is T$_{\rm IGM}<0.5$.
The low luminosities and the extremely large \lya\ EW$_{\rm 0}$ match the case of a Population~III star complex
 made of several dozens stars ($\sim 10^{4}$ \msun) which irradiate a \hii\ region crossing the caustic. While 
 the \lya\ and stellar continuum are among the faintest ever observed at this redshift,
 the continuum and the \lya\ emissions could be  affected by differential magnification, possibly
 biasing the EW$_{\rm 0}$ estimate. The aforementioned tentative HST detection tend to favour a large 
 EW$_{\rm 0}$, making such a faint Pop~III candidate a key target for the James Webb Space Telescope 
 and Extremely Large Telescopes. %
\end{abstract}


\section{Introduction}
Finding and characterising the first galaxies is the next frontier in
observational astronomy. It is thought that
the  Universe  was  initially metal-enriched  by  the  first  generation
of Population  III  (Pop~III)  stars, that could also have
played a key role in cosmic reionisation before the formation of primeval galaxies 
\citep[e.g.,][and references therein]{zack19,wise19,pratika18}.
Late ($z < 7$) Pop III star formation might also
have occurred in pristine regions due to inhomogeneous
metal enrichment of the first galaxies
\citep[][]{torna07,visbal16,ruben11}. 
Given the exceptionally high effective temperatures of Pop~III stars in 
the zero-age main sequence, they emit a large fraction of their luminosity in the
Lyman continuum and have a much harder ionising spectrum
than stars with higher metallicity. The main characteristics of their 
predicted spectral energy distribution (SED) are the presence of
a prominent rest-fame \lya\ (Lyman-alpha) emission line due to the strong ionising
flux up to $\sim 1000-4000$\AA\ rest-frame equivalent width (denoted as EW$_{\rm 0}$, hereafter) 
and significant He recombination line (especially \heii, with
EW$_{\rm 0}$ up to 15-40\AA) due to spectral
hardness, while a clear deficit of all the metal lines is expected.
In particular, \citet{inoue11} suggested the following criteria
for the identification of extremely metal poor or Pop~III galaxies: 
EW$_{\rm 0}$(\lya) $>230$\AA, 
EW$_{\rm 0}$([OIII]5007) $<20$\AA~and EW$_{\rm 0}$(\heii) $>1$\AA,
and prominent Balmer lines like EW$_{\rm 0}$(\ha)$>1900$\AA,
while showing an extremely blue ultraviolet spectral slope 
($\beta \sim -3$, F$(\lambda) \sim \lambda^{\beta}$).

Observations have yielded candidates for Pop~III stellar
populations at high redshift \citep[e.g.,][and references therein]{kashi12,sobral15}, yet
without any definitive detection. These include a controversial $z=6.6$ galaxy  
dubbed CR7 that displays \heii\ emission \citep{sobral19,shi18}. 
Thus, to date, there has not been a confirmed observation of a galaxy dominated by 
the flux of Pop~III stars.
The possibility of observing signatures from very metal poor or Pop~III star clusters 
through gravitational lensing has also been discussed, e.g., \citet{zack15} (see 
also \citealt{shards17}),
including the detection of single Pop~III stars 
with fluxes temporarily magnified to extreme values 
 (with the magnification parameter $\mu \simeq 10^3-10^5$)
during their transit across the caustic of a galaxy cluster. 
Such single$-$star$-$transit events can boost the flux of the
star by $7-12$ mag \citep[][]{wind18}, making such objects 
visible for a limited amount of time even down to intrinsic magnitudes of $35-38$.
Examples of such events detecting single normal stars at $z<2$ have been reported
recently by \citet[][]{rodney18}.

\begin{figure}
\centering
\includegraphics[width=8.0cm]{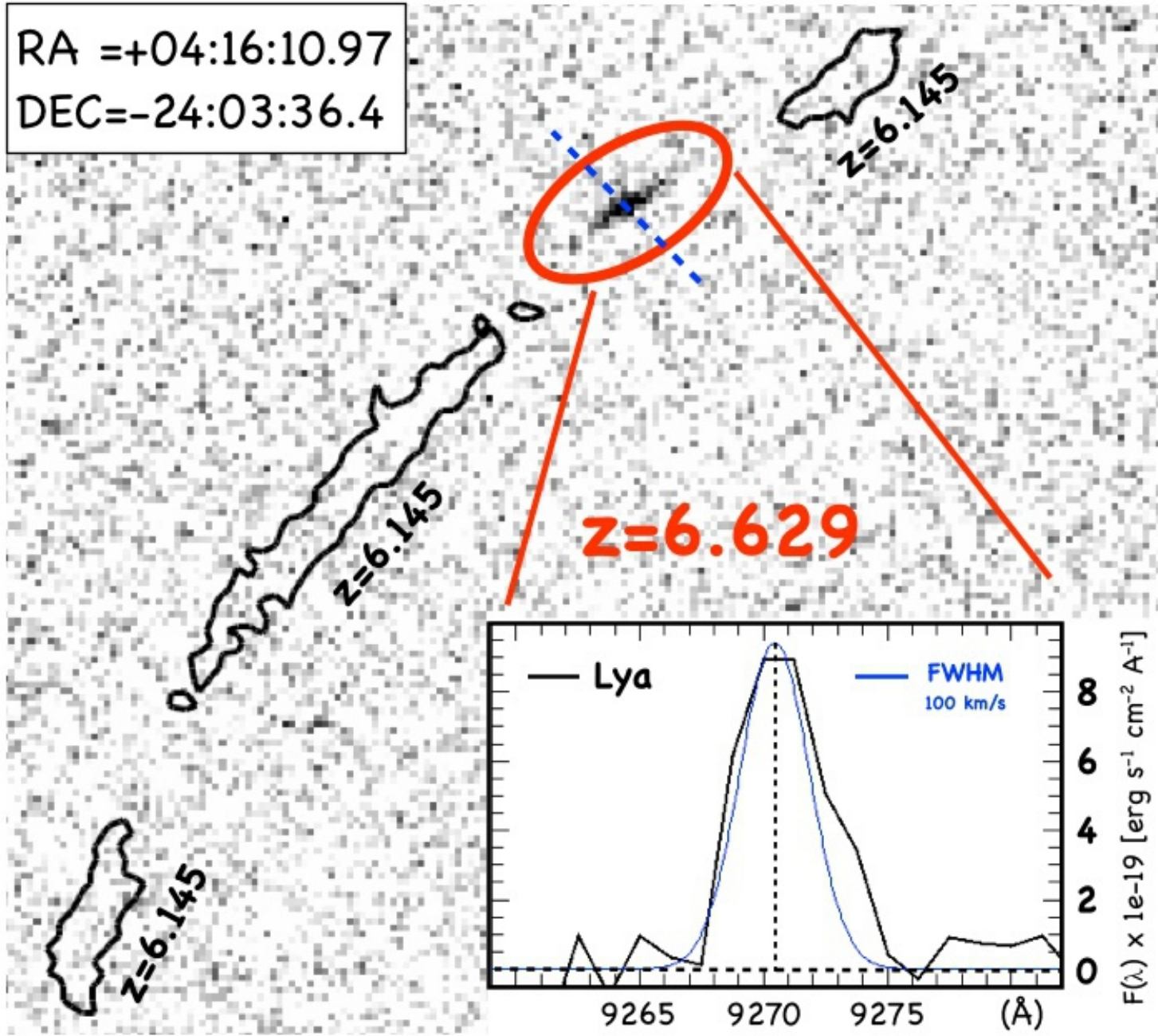}
\caption{A  $30'' \times 30''$ region extracted from the MDLF at the z=6.629 \lya\ wavelength and averaged over $dv=200$ \kms. The indicated \lya\ arclet straddles the critical line (marked with the blue dotted line), close to a highly magnified system already confirmed at z=6.145
and well constrained by the three giant \lya\ arcs (black contours).  The one-dimensional spectrum
of the \lya\ line at z=6.629 is shown in the inset, where the asymmetry towards the red side is evident (a Gaussian with FWHM=100 \kms\ is superimposed with a blue line).}
\label{pano}
\end{figure}

Very low-luminosity emission line galaxies have been identified in Hubble Ultra Deep Field, down to magnitude $30-32$ (M$_{\rm 1500}$ = -15) and S/N $\sim 1-5$  \citep[][]{maseda18}. Strong gravitational lensing allowed us to shed further light on similar low-luminosity objects, providing higher S/N~$\sim 20$ for individual cases \citep[e.g.,][]{vanz17,vanz19}.
In this letter we present an object at z=6.629 showing (1) the faintest \lya\ emission ever detected at $z>6$ crossing the caustic of the Hubble Frontier Field (HFF) galaxy cluster MACS~J0416 \citep{lotz17} and (2) a large \lya\ EW$_{\rm 0}$, potentially implying that extreme stellar populations are present.
We assume a flat cosmology with  $\Omega_{\rm M}$= 0.3, $\Omega_{\Lambda}$= 0.7 and $H_{0} = 70$ km s$^{-1}$ Mpc$^{-1}$.

 \begin{figure*}
\centering
\includegraphics[width=17.5cm]{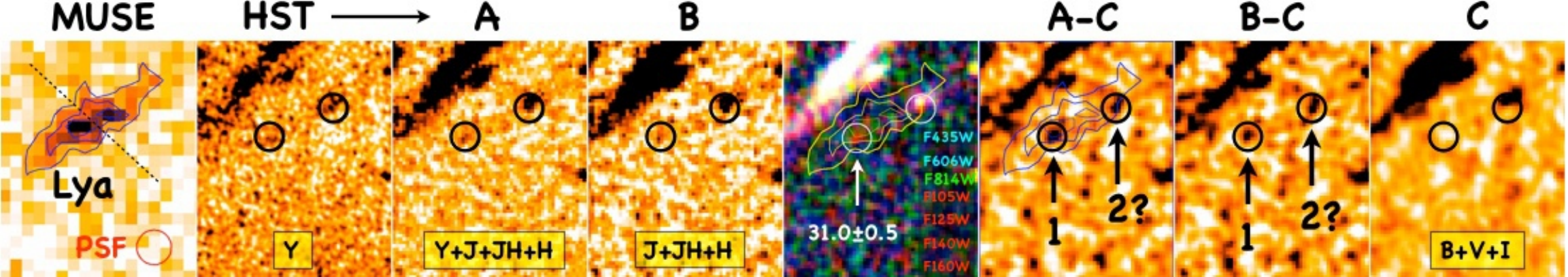}
\caption{From left to right: the MUSE \lya\ emission averaged over $dv = 160$ \kms\ (the 2 and 4 $\sigma$ contours are  shown, with the dotted line marking the critical line); the Y-band HST image; the stacked Y+J+JH+H and J+JH+H  images with indicated the positions of the two possible high-z counterparts (solid circles with diameter $0.4''$); the colour stacked image; the differential stacked HST images highlighting
the two possible counterparts (A-C) and (B-C) after a Gaussian smoothing with $\sigma$=1 pix; the B+V+I stacked image (C).} 
\label{cut}
\end{figure*}

\section{The MUSE Deep Lensed Field: MDLF}

VLT/MUSE \citep{bacon10} deep spectroscopic observations of 17.1 hours of 
integration time in a single pointing have been obtained on the HFF galaxy cluster MACS~J0416 (Prog. {\tt 0100.A-0763(A)}, PI Vanzella).
The data reduction follows the technique described in \citet{cam17}, eventually achieving a PSF with FWHM of $0.6''$ in the final datacube. A more detailed description of the observational campaign of the MDLF and of the data reduction 
will be presented elsewhere. A refined lens model of HFF~J0416 using a new set of confirmed multiple images from the MDLF, will also be presented in a forthcoming paper.
 
\subsection{A \lya\ arc at z=6.629 and the faint HST counterpart}
\label{sect_arc}

Figure~\ref{pano} shows the extended ($3''$) arc from the continuum subtracted narrow band image extracted from the MUSE data cube and the one-dimensional
profile of the emission line at $\lambda=9270.7$\AA, in a region free from OH sky emission lines. 
The arc is detected at S/N=18
with flux $4.4\times 10^{-18}$ \ergscm\ calculated within an elliptical aperture with major and minor axes of $4''$ and  $1.5''$, respectively, and shows an asymmetric profile having an instrumental corrected
FWHM of $98(\pm7)$ \kms. We identify this line as \lya\ at $z=6.629$ for the following reasons:
(1) the weighted skewness $S_{W}$ \citep[as defined by][]{sw06} is $3.4\pm0.7$, in line with the typical 
values observed for asymmetric \lya\ emissions at high-z (it is zero for symmetric shapes, Figure~\ref{pano}); (2)
if it is identified to other typical lines like \oiidoub, \oiiiiv, \oiiiv, \hb, or  \ha,
each of them would imply the presence and detection of additional lines in the same spectrum;
(3) the MUSE spectral 
resolution at $\lambda > 9000$\AA\ is $R\simeq3500$, high enough to resolve the single components of the doublets like \civ, \oiiiuv, \ciiidoub, \oiidoub\ further excluding these lines for identification.
As Figure~\ref{cut} shows, there is no clear detection in the F105W (Y), F125W (J), F140W (JH), and F160W (H) bands in the HFF images, probing the ultraviolet stellar
continuum down to the nominal depth of the 
HFFs \citep[mag $\simeq 29$, at 5$\sigma$ limit][]{lotz17}. We therefore computed the Y+J+JH+H weighted-mean
stacked images (probing $\lambda \sim 1500$\AA), reaching a 1$\sigma$ limit of 31.6 within circular apertures 
of diameter $0.4''$. Such a limit has been derived by inserting  
30 apertures in free regions surrounding the source position and computing the standard deviation among them
(the A-PHOT tool has be used for this task, \citealt{merlin19}). As discussed in the next section, there is a configuration in which we expect 
the presence of two very close multiple images near the \lya\ arc. Indeed, there is a possible
detection at S/N~$\sim2$ lying within the arclet indicated as 1 in Figure~\ref{cut}, 
with $m_{\rm 1500} \simeq 31$ and showing a photometric drop in the F435W+F606W+F814W image.
The same test has been performed adopting an elliptical aperture oriented along the arc, with semi-axis $0.7''$ and 
$0.2''$, and no signal has been detected down to $m_{\rm 1500} \simeq 30.85$ at $1\sigma$.
We expect a second nearby image with similar magnification that, however,
is contaminated by a foreground object clearly detected in the blue bands. 
While image 1 could be the HST counterpart, a tentative second image detection marked as 2 is shown in Figure~\ref{cut}.


\section{The \lya\ emission is on the caustic}
\label{sect_scheme}

The \lya~arc lies in a well known region of the galaxy cluster where \citet{vanz17,vanz19} already discussed another star-forming complex system at z=6.145 showing several multiple images identified in deep HST data, producing three clear \lya\ arcs in the MUSE observations (Figure~\ref{pano}). The presence  of such a system adds valuable constraints for the  case studied in this work. 
In fact, any detection at $z>6$ in the region where the  z=6.629 arc lies, would produce multiple images as in the case observed at z=6.145,
unless such images are so close to merge into a single spatially unresolved mildly elongated arc. It is exactly the case for the z=6.629 arclet discussed here: the absence of two distinct images (Figure~\ref{pano} and \ref{cut}) implies that the \lya\ arc straddles the critical line and is indeed the result of two spatially unresolved \lya~images, generated by a \lya~emitting region lying on the corresponding caustic. 

\begin{figure}
\centering
\includegraphics[width=8.0cm]{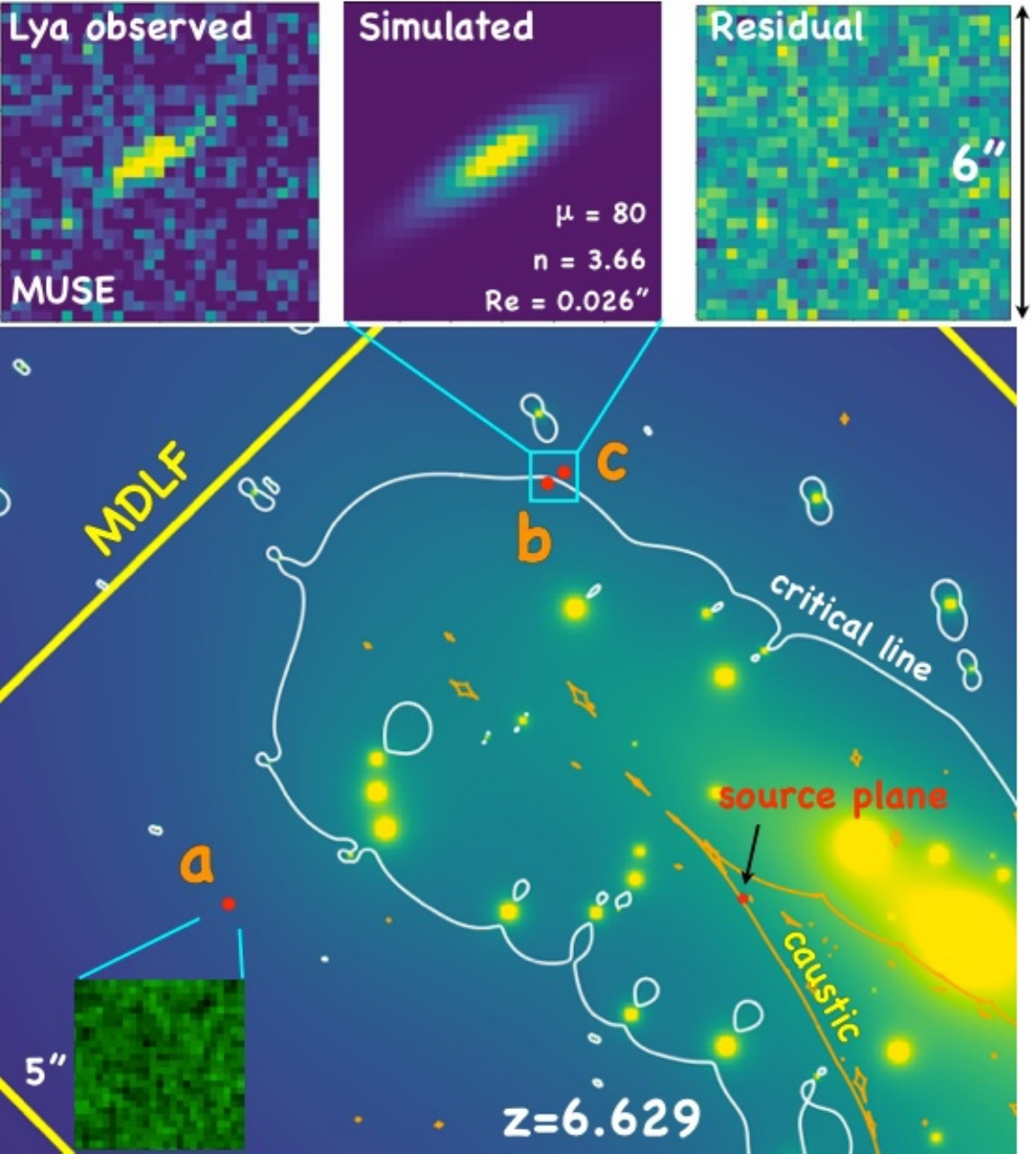}
\caption{The tangential caustic (orange) and the critical line (white) are superimposed to the (false-colour) image of the galaxy cluster members used for the strong lensing model. The observed \lya\ arc is well reproduced when the source is very close to the caustic (black arrow). Three multiple images are predicted (marked as a, b and c): images b,c merge into a single small arc, whereas image a is not detected (bottom-left inset). In the top panels, the observed, simulated and residuals \lya\ images are shown for the best fit values of $n$, $R_e$ and $\mu=\mu(a)+\mu(b)$.}
\label{simul}
\end{figure}
 
  \subsection{Simulating the caustic crossing}
\label{simula}

In order to perform a quantitative estimate of the magnification of the \lya\ emission, we use dedicated simulations with a customised version of the software {\tt SkyLens} \citep[e.g.][]{max2010,pizas19}.
The method will be extensively described in a future paper. In short, we perform the following steps. We start from the assumption that the source of the \lya\ emission can be described by a single, circularly symmetric S\'ersic surface brightness profile. The profile is characterised by the S\'ersic index $n$ and by the effective radius $R_e$. We ray-trace $2000\times 2000$ light rays through a squared region containing the \lya\ emission (cyan square in Figure.~\ref{simul}, whose size is $\sim 6.2''\times 6.2''$) and propagate them towards the source plane at redshift $z_s=6.629$, accounting for the deflections induced by the lensing cluster. In this work, we use the strong lensing model described in \citet{cam17}\footnote{The LensTool model is published in the HFF lens model format at {\it http://www.fe.infn.it/astro/lensing/}}. 
On the source plane, the arrival positions of the light rays are used to sample the brightness distribution of the source, which is then mapped onto the image plane. By performing this operation, we reconstruct the lensed image, which is subsequently convolved with a Gaussian PSF with a FWHM of $0.6"$. The resulting convolved image is first rebinned at the same resolution of the MUSE image ($0.2''$/pixel) and then  compared to the observed \lya\ arclet. We use the python package {\tt LMfit} to perform a Non-linear Least Square Minimisation of a cost function defined as the squared difference between the simulated and the observed images. 
As shown in the upper panels of Figure.~\ref{simul}, the source model, despite its simplicity, reproduces very well the observed arclet morphology. 
The two parameters $n$ and $R_e$ are quite degenerate. For a S\'ersic index in the range $n=1-4$, the best fit  effectitve radius $R_e$ varies between $0.012''-0.027''$. The best fit solution corresponds to $n=3.66$ and $R_e=0.026''$. The latest is equivalent to 140 pc on the source plane. In all cases, the model suggests that the  \lya\ emitting region partially overlaps with the caustic, meaning that only part of the source is reproduced twice in the arclet. 

Having obtained a model for the \lya\  source, we can  estimate the magnification of the arclet
by comparing the lensed and the intrinsic fluxes of the source after integrating the surface brightness over the 
region used for ray-tracing. We estimate that the total magnification of the \lya\ arclet (images b and c) 
is $\mu(b+c) = \mu(b)+\mu(c) \simeq 80$, implying that the de-lensed \lya\ flux of the arclet is $\sim 5.5 \times 10^{-20}$ \ergscm. Keeping $n\sim1$ (exponential profile) in our fitting procedure the magnification increases to $\mu(b+c) \sim 110$. 

It is well known that such magnification estimate is prone to systematic errors due to the uncertainties of the lens model in regions where the magnification gradients are very strong  \citep[e.g.,][]{max17}. In order to circumvent this limitation, we can make use of the counter image {\rm a}, which is predicted to form much farther from the critical lines (Figure.~\ref{simul}). Using our best fit source determined above, we estimate that $\mu(a) \sim 4.5$
with an uncertainty smaller than $20\%$.  We also find that, because of the much shallower magnification gradient in this region, the estimate is quite insensitive to the properties of the source. Indeed, $\mu(a)$ would change by $\sim 5\%$ ($\mu(a) \sim 4.3$) by adopting a point source approximation. 
The image $a$ is not detected at the depth of the MDLF down to $1\sigma$ \lya\ flux limit $f_{a,lim}$ ($\sim 2.4 \times 10^{-19}$ \ergscm, for a point-like source, consistently to \citealt{inami17}). This sets a lower limit $\mu(b+c)\gtrsim  \mu(a)f_{bc}/ f_{a,lim}$, where the observed ratio is $f_{bc}/ f_{a,lim} \simeq 18$. Therefore, we can conclude that $\mu(b+c)\gtrsim 80$, in keeping with the fitting procedure ($\mu = 80-110$). 

\subsection{A large \lya\ EW}
\label{EW}

The computation of the EW of the \lya\ line (L) requires an estimate of the underlying stellar continuum (S), taking into account that magnifications associated to S ($\mu_S$) and L ($\mu_L$) might differ. A general expression for EW$_{\rm 0}$ is:
\begin{equation}
    EW_{0} =  \frac{1}{(1+z)} \frac{\mu_{S}}{\mu_L} \frac{f(\lya)}{F_{\lambda}(UV) \times T_{IGM}(\lya)} \;,
\label{eq1}
\end{equation}
where $f$(\lya) is the \lya\ flux ($5.5\times10^{-20}$ \ergscm), $F_{\lambda}$(UV) is the ultraviolet continuum
at the \lya\ wavelength for which we assume a value $<1.27\times10^{-23}$ erg~s$^{-1}$~cm$^{-2}$\AA$^{-1}$, corresponding to
 $m_{1500} \gtrsim 35$ ($=$ 31+2.5Log$_{10}(\mu)$, $\mu \gtrsim 40$). Given the large uncertainties on the HST detection (S/N~$\sim2$, $\sigma_m \sim 0.5$), any assumption on
the ultraviolet slope $\beta$ would not be significant. Indeed, a slope $\beta = -2.5$ ($-3$) would imply
a magnitude difference of $m_{1216}-m_{1500}=-0.11~(-0.23)$.
$T_{\rm IGM}$(\lya) is the transmission of the intergalactic medium for \lya\ photons (see below, and Table~\ref{infos}). 
We identify two scenarios:

\noindent $\bullet$ if we assume that {\em L} and {\em S} have the same intrinsic size and brightness profile, then  $\mu_S \simeq \mu_L$ and the $EW_{\rm 0}$ is independent on the lens model. Thus, under this assumption, 
a lower limit on $EW_{\rm 0}$ can be found by using the very low significance detection (if not the non-detection) of the UV continuum, $m_{\rm 1500} \gtrsim 35$. Combined with the de-lensed \lya\ flux of $5.5\times 10^{-20}$ \ergscm\ leads to the result that $EW_{\rm 0} > 564$\AA, in the case T$_{\rm IGM}$(\lya)=1.0. The low significance of the HST detection prevents us from verifying whether S is extended as the arclet.

\noindent $\bullet$ if, in contrast, we assume that the size of {\em S} is smaller than that of {\em L}, as it might be the case when {\em S} is embedded and generates the \hii\ region \citep[e.g.,][]{steidel11}, then we expect that  $\mu_S > \mu_L$ and consequently EW$_{\rm 0}$ would be even larger than in the previous case.

There is still the possibility that $S$ is located outside the lens caustic. In this scenario no  continuum flux is expected near the \lya\ arclet. We could only use the non-detection in the HST data of image $a$ ($m>31.6$ at $1\sigma$), where $S$ is certainly present, to set an upper limit of $F_{\lambda}$(UV). Using the fact that $\mu(a) \sim 4.5$, the de-lensed magnitude limit is $m>33.2$. Combined with the lower limit of the magnification of $L$ ($\mu_{L}\gtrsim 80$), we obtain that $EW_{\rm 0} \gtrsim 110$\AA\ in the case T$_{\rm IGM}$(\lya)=1.0. Note, however, that the marginal detection of image 1 in the stacked HST images (see Figure~\ref{cut}), which could be the image of $S$, seems to disfavour this scenario. 

It is now worth discussing the IGM transmission T$_{\rm IGM}$(\lya) which depends on {\it both} the ``intrinsic'' (pre-IGM) \lya\ spectrum emerging {\it and} the IGM properties. Due to the resonant absorption of the neutral or partially neutral IGM combined with cosmological inflow, the \lya\ spectrum blueward of $v_{\rm cutoff} \lesssim 200$\kms\ is absorbed \citep[][]{dijkstra07}. Thus, the more asymmetric the intrinsic \lya\ is towards the blue or the lower the red peak offset from systemic, the lower T$_{\rm IGM}$. 
All this implies that T$_{\rm IGM}$ is highly uncertain and estimates reach from tens of percent (e.g., T$_{\rm IGM}=0.20^{+0.12}_{-0.15}$ at $z=6.6$, \citealt{laursen11}) to values approaching unity for an intrinsic single read peak with offset $> 300$ \kms. However, since the \lya\ spectrum presented here is very narrow and asymmetric (cf. Sect.~\ref{sect_arc}) it is likely that either a significant part of the red peak has been removed from the IGM (in case of an intrinsic spectrum with a large offset, and thus, large width), or an intrinsic blue component existed (in the case of a small intrinsic offset). Both cases would imply a significant absorption of the IGM, and hence, T$_{\rm IGM} < 1$. Knowledge of the systemic redshift, through, e.g., \ha\ information would be helpful in reconstructing the intrinsic \lya\ line, and thus, to constrain T$_{\rm IGM}$ more qualitatively. This could, furthermore, rule out the radiative transfer effects as an origin of the large EW -- which
we already deem unlikely due to the asymmetry of the observed line.

We conclude that, even assuming T$_{\rm IGM}$(\lya)=1, a still quite extreme EW$_{\rm 0}$(\lya)~$>550$\AA\ emerges from a region crossing the caustic,
that can easily approach (or exceed) 1000\AA\ assuming a more plausible T$_{\rm IGM}$(\lya)~$<1.0$.
Though not totally excluded, we do not consider in this work the possibilities that the large EW(\lya) originates from a very faint AGN (e.g., with BH mass of $10^{2-3}$ \msun, \citealt{fan12}) or a multiphase scattering medium \citep[e.g.,][]{neufeld91}.

\section{Candidate Pop~III stars}

The predicted EW$_{\rm 0}(\lya)$ for metal free stellar populations exceeds 400\AA\ and it goes up to a few thousands rest-frame \citep[][]{inoue11,schaerer13},
and is observable if neighbouring sources (either Pop~III or Pop~II stars) have already contributed toward ionising a local 
bubble \citep[][]{sobral15}. The large EW$_{\rm 0}(\lya)$ value reported in this work opens for a possible dominant contribution by 
extremely metal poor stars.
It is interesting to calculate how many Pop~III stars are needed to reproduce both the observed 
M$_{\rm 1500}$ and L(\lya): 

\noindent $\bullet$ {\it UV continuum}: the apparent magnitudes at 1500\AA~rest-frame at $z=7$ for  Pop~III star at ZAMS with
masses $1-1000$ \msun\ are reported by 
\citet{wind18}. In particular, stars with masses of 100, 300, and 1000\msun\ have magnitudes $m_{\rm 1500}=40.08$, $38.64$, and $37.44$, respectively, neglecting dust attenuation. 
Adopting $m_{\rm 1500} \gtrsim 35$ (M$_{\rm UV} \gtrsim -11.9$, Sect.~\ref{EW}) 
and assuming for simplicity the same masses for all stars, the number of Pop~III stars required to reproduce the intrinsic UV flux corresponding to M$_{\rm UV}$ amounts to N(M$_{\rm UV}$) = 10, 30, and 110 for stellar masses of 1000, 300 and 100\msun.

\noindent $\bullet$ {\it \lya\ emission}: 
 \citet[][]{masribas16} provide the photon flux Q(\hi)[s$^{-1}$] for different Pop~III ZAMS stars and
the conversion to L(\lya) luminosity considering case-B departure, stochastic sampling of the Salpeter and Top-Heavy IMFs
and zero escape fraction of the Lyman continuum radiation (note, however, that if a fraction of the ionising radiation escapes, the emerging \lya\ would be dimmed linearly by the same factor, e.g., \citealt{schaerer13}). 
We perform the calculation as above, assuming again the same mass for all stars (no boosting from the stochastic sampling of the IMF is considered).
The \lya\ luminosity emerging from Pop-III stars of mass 1000, 300, and 100 \msun\ are $3.20\times 10^{40}$, $8.11\times 10^{39}$, and $1.72\times 10^{39}$ \ergs. 
Under the assumption that N(\lya) = N(M$_{\rm UV})$ (being \lya\ and M$_{\rm UV}$ referring to the same star complex),
 the resulting T$_{\rm IGM}$(\lya) are 0.09, 0.11, and 0.16, respectively. 
These values double if the case-B is assumed, i.e., the predicted L(\lya) is about a factor two fainter \citep[][]{masribas16}.
With such values of T$_{\rm IGM}$(\lya) the resulting EW$_{\rm 0}$(\lya) ranges between 4000-1500\AA, for the three classes of Pop~III stellar masses.
 
\begin{table}
\footnotesize
\caption{Properties of the \lya\ emitter in the source plane.}
\begin{tabular}{l r}
\hline
\lya\ [\ergscm] & $ 5.5\times10^{-20}$\\
\lya\ [\ergs] & $ 2.8\times10^{40}$ \\
EW$_{\rm 0}$(\lya) [\AA] (T$_{\rm IGM}<0.5$) & $> 1120$\\
M$_{\rm 1500}(m_{\rm 1500})$~($2\sigma$) & $\gtrsim -11.9~(\gtrsim 35)$ \\
$R_e$ \lya\ region [pc] & $< 150$\\
Magnification [$\mu(b)=\mu(c)$] & $\mu(b) + \mu(c) \gtrsim 80$; $\mu(a)\simeq 4.5$ \\
\hline
\hline 
\end{tabular}
\label{infos}
\end{table}

\noindent Future facilities are necessary 
to make a significant step forward. First, only the James Webb Space Telescope will access the optical rest-frame looking for the possible deficit of metals and the expected enormous Balmer emissions \citep[e.g.,][]{inoue11}, eventually gaining in depth with respect HST imaging.
The next generation of Extremely Large Telescopes will also investigate the currently vague stellar component S by performing very deep imaging, while spectroscopy will address the deficiency of high-ionisation metal lines and the possible key \heii\ emission.
The intrinsic \heii/\lya\ line ratio predicted for Pop~III spans the range 0.01-0.10 \citep[e.g.,][]{schaerer13,masribas16}, implying the
expected flux of \heii\ would be $1.1 \times  (10^{-21} - 10^{-20})$/T$_{\rm IGM}$(\lya) \ergscm, clearly requiring an ELT-like telescope  or an 8-10m class telescope in the most optimistic cases (T$_{\rm IGM}$(\lya) $\ll 1$).

\section*{Acknowledgments}
We thank the anonymous referee for a constructive report.
We thank Ll. Mas-Ribas and D. Schaerer for very stimulating discussions.
This work is supported by PRIN-MIUR 2017 WSCC32.
We  acknowledge funding from the INAF
main-stream (1.05.01.86.31). KC and GBC acknowledge funding from the ERC through the award of the Consolidator Grant ID 681627-BUILDUP.
This work was supported in part by the NSF grant: COLLABORATIVE RESEARCH: The Final Frontier: Spectroscopic Probes of Galaxies at the Epoch of Reionization
(AST-1815458, AST-1810822). MG acknowledges funding through HST-HF2- 51409.

\end{document}